# Model Based Ceramic tile inspection using Discrete Wavelet Transform and Euclidean Distance

Samir Elmougy[1], Ibrahim El-Henawy[2], and Ahmed El-Azab[3]

[1]Dept. of Computer Science, College of Computer and Information Sciences, King Saud Univ., Riyadh 11543, Saudi Arabia

[1]Dept. of Computer Science, Faculty of Computer and Information Sciences, Zagazig University, Zagazig, Egypt

[3]Dept. of Computer Science, Misr for Engineering and Technology (MET) Academy, Mansoura, Egypt

*Abstract*— **Visual inspection of industrial products is used to determine the control quality for these products. This paper deals with the problem of visual inspection of ceramic tiles industry using Wavelet Transform. The third level the coefficients of two dimensions Haar Discrete Wavelet Transform (HDWT) is used in this paper to process the images and feature extraction. The proposed algorithm consists of two main phases. The first phase is to compute the wavelet transform for an image free of defects which known as reference image, and the image to be inspected which known as test image. The second phase is used to decide whether the tested image is defected or not using the Euclidean distance similarity measure. The experimentation results of the proposed algorithm give 97% for correct detection of ceramic defects.**

## I. INTRODUCTION

Visual inspection of industrial product is one of the main important phases in many industries. It is used to determine the quality for the control for some products such as wood [1], textile [2], leather [3], steel [4], Printed Circuit Board (PCB) [5] and ceramic tiles [6]. In realty, ceramic tiles industry has hazardous, high polluted and unhealthy environment [7]. The inspection in this industry was usually made by humans so it is important to use mechanical technology to instead of standard humans to keep them healthy. A large variety of fast and different algorithms for object detection and recognition has been studied during the last decade by the computer vision community [8]. These algorithms can be divided into main approaches such as statistical, structured, filter based and model based [9]. In this work, a model based approach using DWT and Euclidean distance is introduced.

The earliest based model for visual inspection of ceramic tiles was carried by image difference operation (pixel-by-pixel comparison like XOR logic operator) [5]. Although this model has a good recognition, its operation costs too much processing time, and requires much memory more over the alignment of the tested image which should be identical to the reference image. Figure (1) depicts the image difference operation on ceramic tile.

In this paper, the proposed model is based on discrete wavelet transform because wavelet transform usually lead to a better image modeling, a better image encoding (this is the reason why wavelet is used as one of the best compression methodologies), and a better texture modeling.

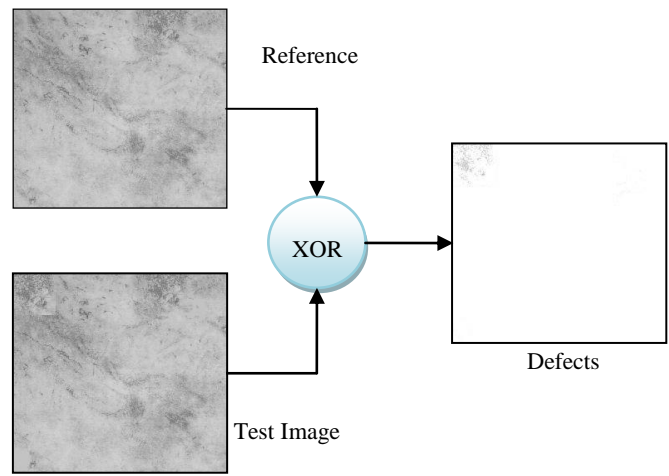

Figure 1. Image difference operation of ceramic tile.

The rest of the paper is organized as follows. Overview of Wavelet processing is given in Section 2 followed by the Continuous Wavelet Transform (CWT), 1-D Discrete Wavelet Transform and 2-D Discrete Wavelet Transform in Subsections 2.1, 2.2 and 2.3 respectively. The proposed algorithm and its results are shown in Section 3. LVQ neural network structure and its algorithm explained in Section 4. Finally, conclusion and future work are discussed in Section 5.

## II. WAVELET PROCESSING

Because the frequency contents of signals are very important, transforms are usually used. The earliest well known transform is Fourier transform which is a mathematical technique for transforming our view of the signal from time domain to frequency domain. Fourier transform breaks down the signal constituents into sinusoids of different frequencies. However, Fourier transform comes with serious shortage that is the lost of time information which mean it is impossible to tell when a particular event take place [10]. This shortage vanishes with using wavelet transform. A shifted version of the original signal is called *mother* wavelet which it is a wave form effectively a limited duration and its average value is zero. The most well known wavelets are Haar. Figure (2) depicts some types of these wavelets [11].

### A. Continuous Wavelet Transform

The Continuous Wavelet Transform (CWT) given in Equation (1), where $x(t)$ is the signal to be analyzed, and $\psi(t)$ is the mother wavelet or the basis function which it must be integrated to zero as given in Equation (2). All the wavelet functions used in the transformation are derived from the mother wavelet (Figure 2) through translation (shifting) and scaling (dilation or compression).

$$X_{WT}(\tau,S) = \frac{1}{\sqrt{|S|}} \int X(\tau) . \Psi^* \left(\frac{t-\tau}{S}\right) dt \qquad (1)$$

$$\int \Psi(t) dt = 0 \qquad (2)$$

Note that $\tau$ and $S$ are real numbers representing translation and scaling parameters respectively. The translation parameter $\tau$ relates to the location of the wavelet function as it is shifted through the signal. Thus, it corresponds to the time information in the Wavelet Transform. The scale parameter $S$ shows either dilates (expands) or compresses a signal. Scaling parameters are calculated as the inverse of frequency [12].

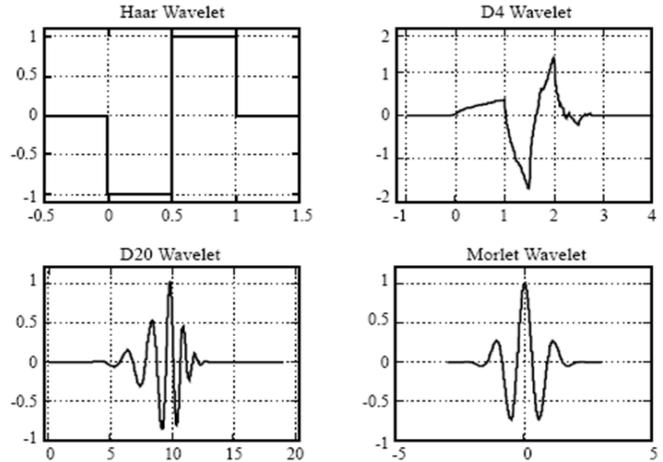

Figure 2. Most popular Wavelets.

### 1) 1-D Discrete Wavelet Transform

The CWT calculates coefficients at every scale which leads to need much time and awful lot amount of data. If scales and positions are selected based on powers of two, analysis will be much more efficient and accurate. This type of selection is called dyadic scales and positions. This analysis can be produced from the Discrete Wavelet Transform (DWT) [7]. DWT is used to decompose (analyze) the signal into approximation and detail called coefficients. Approximation coefficients represent the high scale (low frequency) components of the signal as if it is a low pass filter. Detail coefficients represent the low scale (high frequency) components of the signal as if it is a high pass filter. Given a signal $S$ of size $N$, downsampling the approximation coefficients ($cA$) is given by $N/2$ and the detail coefficients ($cD$) is given by $N/2$ (Fig. 3).

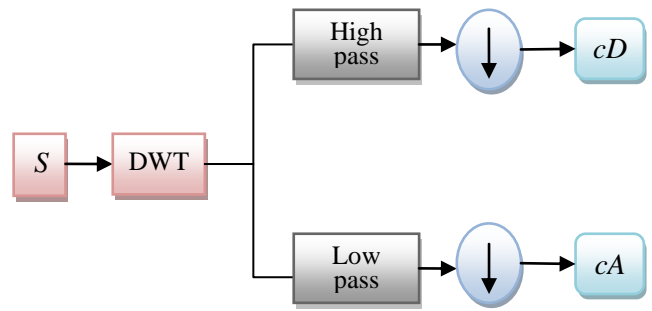

Figure 3. 1-D discrete wavelet transform

The decomposition process of DWT can be iterated to the first time approximation coefficients $cA1$ resulting second detail coefficients $cD2$ and second approximation coefficients $cA2$ which can be decomposed again. This process is known as the Wavelet decomposition tree (Fig. 4-a) and its inverse operation of decomposition is called reconstruction, or synthesis. Reconstruction is used to retrieve the signal back from wavelet coefficients without lose of information. The reconstruction of the signal is done using Inverse Discrete Wavelet Transform (IDWT) operation (Fig. 4-b).

*2) 2-D Discrete Wavelet Transform*

Discrete Wavelet Transform (DWT) is not only applied to 1-D signals, but also applied to two dimensional matrixes applied images. Each element in the matrix represents the intensity of gray color in the image. The computation of the wavelet transform of image is applied as a successive convolution by a filter of row/column followed by a column/row. The results of DWT on image are four coefficients matrices [5].

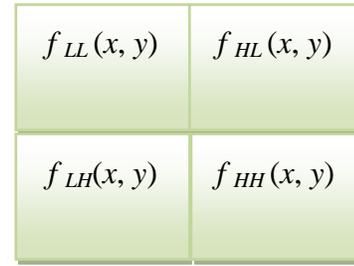

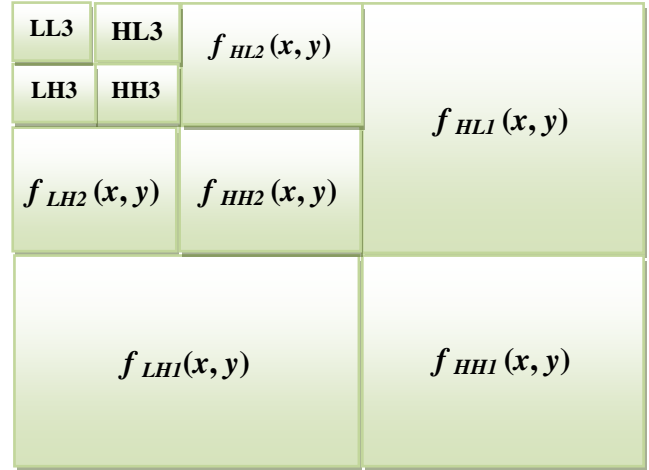

(b)
Figure (5). a) The concept of first level DWT, b) The concept of third level DWT

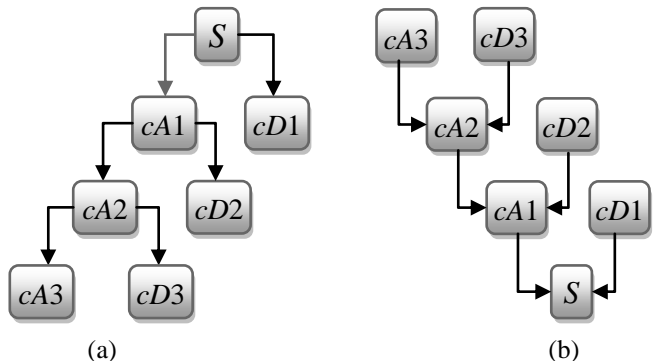

Figure (4). a) 1-D DWT decomposition tree, b) 1-D DWT reconstruction tree

Given image $f(x, y)$, the 2-D wavelet analysis operation consists of filtering and down-sampling horizontally using a 1-D low pass filter $L$ and a high pass filter to each row in the image $f(x, y)$, and produces the coefficient matrices $f_L(x, y)$ and $f_H(x, y)$. Vertically, filtering and down-sampling follow using the low pass and high pass filters $L$ and $H$ to each column in $f_L(x, y)$ and $f_H(x, y)$. This produces 4 sub-images $f_{LL}(x, y)$, $f_{LH}(x, y)$, $f_{HL}(x, y)$ and $f_{HH}(x, y)$ for one level of decomposition. $f_{LL}(x, y)$ is a smooth sub-image, which represents the approximation of the image. $f_{LH}(x, y)$, $f_{HL}(x, y)$, and $f_{HH}(x, y)$ are detail sub-images which represent the horizontal, vertical and diagonal directions of the image respectively [14]. As mentioned before, DWT can be applied again to the approximation $f_{LL}(x, y)$ where the resulted coefficients matrix of approximation and details of DWT determined by the level $k$ of decomposition using the relation $3k+1$. Fig. (5-a) and (5-b) show the first and third level concepts of DWT for image $f(x, y)$.

## III. PROPOSED ALGORITHM

There are four main steps in visual process. These steps are image capturing, preprocessing, feature extraction and classification. The third and fourth steps are the most important process. In this paper, the third level of the discrete Haar wavelet transform as a feature extraction of ceramic tiles' images is used. Haar is selected to be used as a mother wavelet because it has the smallest filter length so the processing time can be minimized [5].

The classification is done with statistical similarity measure using Euclidean distance between the third level approximation of both the reference image and test image. Euclidean distance is the square root for the summation of squared differences between the two approximated images; reference and test. The Euclidean distance $d(i, j)$ between the reference ($r$) and test ($t$) images both of size n × n is represented in Eq. (3).

$$d(i,j) = \sqrt{|r_{0,0} - t_{0,0}|^2 + |r_{1,1} - t_{1,1}|^2 + ... + |r_{m,n} - t_{m,n}|^2} \quad (3)$$

The value of $d(i, j)$ is always greater than or equal to zero. The test image is not defected if the $d(i, j) = 0$ and is defective $d(i, j) > 0$. The low value of $d(i, j)$ means less defects in the test image and more defects otherwise. The proposed algorithm for ceramic tile inspection based on DWT is shown in Fig. (6).

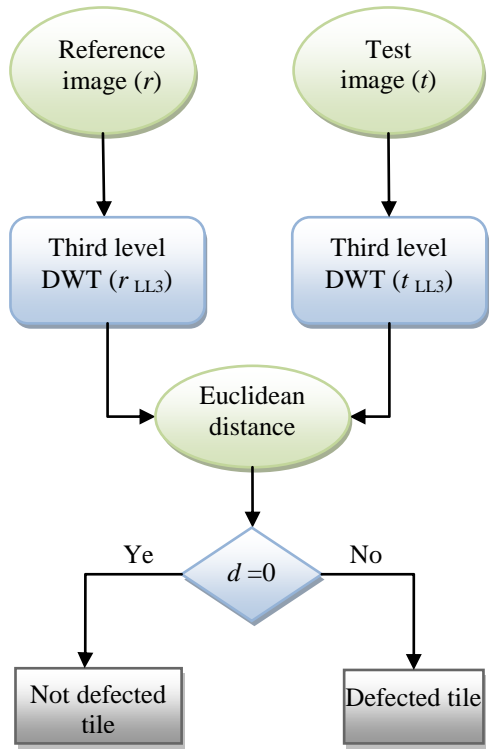

Figure (6). Model based ceramic tile inspection algorithm using DWT

## IV. EXPERIMENTAL RESULTS

The simulation of our algorithm is carried out on 85 images of 256 ×256 pixels resolution using image processing, wavelet, and statistics toolboxes of Matlab software running on Pentium IV PC of processor 1.8 GHz and 512 MB of RAM. Fig. (7-a), (7-b) and (7-c) show the result of DWT on reference, test, and the defects found in the test ceramic tile. The classification accuracy (CA) computed for this algorithm using Eq. (4) is 97%.

$$CA = \frac{No.of \cdot correct \cdot tiles}{Total \cdot No.of \cdot tiles} \times 100 \qquad (4)$$

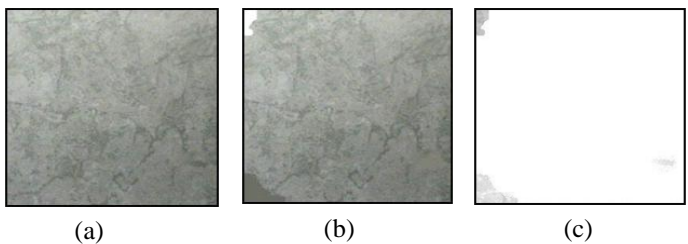

(a)      (b)      (c)

Figure (7). a) The third DWT of reference image, b) The third DWT of test image c) defects of b.

## V. CONCLUSION AND FUTURE WORK

Using Discrete Wavelet Transform (DWT) for image processing and feature extraction with Euclidean distance gives subtle results for the inspection of ceramic tiles surfaces. Also the running time of the proposed algorithm is highly acceptable. This is because the third level DWT is decreasing the size of the image to eighth which reduces the similarity measurement step.

As a future work, we tend to combine Discrete Wavelet Transform (DWT) and the co-occurrence matrix on colored ceramic tiles rather than gray ones. We will try to minimize the running time as it will be greater in processing colored image rather than gray images.